\newcommand{\myname}{Geoff Boeing}
\newcommand{\myemail}{g.boeing@northeastern.edu}
\newcommand{\myaffiliation}{School of Public Policy and Urban Affairs\\Northeastern University}
\newcommand{\paperdate}{August 2018}
\newcommand{\papertitle}{A Multi-Scale Analysis of 27,000 Urban Street Networks: Every US City, Town, Urbanized Area, and Zillow Neighborhood}
\newcommand{\papercitation}{Boeing, G. 2018. \papertitle. \emph{Environment and Planning B: Urban Analytics and City Science}. \url{https://doi.org/10.1177/2399808318784595}}
\newcommand{\paperkeywords}{City Planning, Transportation, Data Science, Street Networks, Python, OpenStreetMap, Graph Theory, Urban Form, Urban Design}

\RequirePackage[l2tabu,orthodox]{nag}
\documentclass[11pt]{article}

\usepackage[T1]{fontenc}
\usepackage[utf8]{inputenc}
\usepackage{crimson}
\usepackage{helvet}

\usepackage[strict,autostyle]{csquotes}
\usepackage[USenglish]{babel}
\usepackage{microtype}

\usepackage{authblk}
\usepackage{booktabs}
\usepackage[final]{draftwatermark}
\usepackage{endnotes}
\usepackage{geometry}
\usepackage{graphicx}
\usepackage{hyperref}
\usepackage{multicol}
\usepackage{natbib}
\usepackage{rotating}
\usepackage{setspace}
\usepackage{titlesec}
\usepackage{url}

\graphicspath{{./figures/}}

\titleformat{\section}{\normalfont\sffamily\large\bfseries\color{black}}{\thesection.}{0.3em}{}
\titleformat{\subsection}{\normalfont\sffamily\small\bfseries\color{black}}{\thesubsection.}{0.3em}{}

\hypersetup{
    pdfauthor={\myname},
    pdftitle={\papertitle},
    pdfsubject={\papertitle},
    pdfkeywords={\paperkeywords},
    pdffitwindow=true,         
    breaklinks=true,           
    hidelinks=true             
}

\SetWatermarkText{DRAFT}
\SetWatermarkScale{1.3}
\SetWatermarkLightness{0.9}

\begin{document}
	
\title{\papertitle\footnote{Cite as: \papercitation}}
\date{\paperdate}
\author[]{\myname \thanks{Email: \href{mailto:\myemail}{\myemail}}}
\affil[]{\myaffiliation}

\maketitle

\begin{abstract}
OpenStreetMap offers a valuable source of worldwide geospatial data useful to urban researchers. This study uses the OSMnx software to automatically download and analyze 27,000 US street networks from OpenStreetMap at metropolitan, municipal, and neighborhood scales---namely, every US city and town, census urbanized area, and Zillow-defined neighborhood. It presents empirical findings on US urban form and street network characteristics, emphasizing measures relevant to graph theory, transportation, urban design, and morphology such as structure, connectedness, density, centrality, and resilience. In the past, street network data acquisition and processing have been challenging and ad hoc. This study illustrates the use of OSMnx and OpenStreetMap to consistently conduct street network analysis with extremely large sample sizes, with clearly defined network definitions and extents for reproducibility, and using nonplanar, directed graphs. These street networks and measures data have been shared in a public repository for other researchers to use.
\end{abstract}

\vspace{0.25cm}

\begin{multicols}{2}

\section{Introduction}

On May 20, 1862, Abraham Lincoln signed the Homestead Act into law, making land across the United States' Midwest and Great Plains available for free to applicants \citep{porterfield_homestead_2005}. Under its auspices over the next 70 years, the federal government distributed 10\% of the entire US landmass to private owners in the form of 1.6 million homesteads \citep{lee_kansas_1979, sherraden_inclusion_2005}. New towns with gridiron street networks sprang up rapidly across the Great Plains and Midwest, due to both the prevailing urban design paradigm of the day and the standardized rectilinear town plats used repeatedly to lay out instant new cities \citep{southworth_streets_1997}. Through path dependence, the spatial signatures of these land use laws, design paradigms, and planning instruments can still be seen today in these cities' urban forms and street networks. Cross-sectional analysis of American urban form can reveal these artifacts and histories through street networks at metropolitan, municipal, and neighborhood scales.

Network analysis is a natural approach to the study of cities as complex systems \citep{masucci_random_2009}. The empirical literature on street networks is growing ever richer, but suffers from some limitations---discussed in detail in \citep{boeing_osmnx:_2017} and summarized here. First, sample sizes tend to be fairly small due to data availability, gathering, and processing constraints: most studies in this literature that conduct topological or metric analyses tend to have sample sizes ranging around 10 to 50 networks \citep{buhl_topological_2006, cardillo_structural_2006, barthelemy_modeling_2008, strano_urban_2013}, which may limit the generalizability and interpretability of findings. Second, reproducibility has been difficult when the dozens of decisions that go into analysis---such as spatial extents, topological simplification and correction, definitions of nodes and edges, etc.---are ad hoc or only partly reported \citep[e.g.,][]{porta_network_2006, strano_urban_2013}. Third, and related to the first two, studies frequently oversimplify to planar or undirected primal graphs for tractability \citep[e.g.,][]{buhl_topological_2006, cardillo_structural_2006, barthelemy_modeling_2008, masucci_random_2009}, or use dual graphs despite the loss of geographic, metric information \citep{batty_network_2005, jiang_integration_2002, ratti_space_2004, crucitti_centrality_2006, crucitti_centrality_2006-1}. 

This study addresses these limitations by conducting a morphological analysis of urban street networks at multiple scales, with large sample sizes, with clearly defined network definitions and extents for reproducibility, and using nonplanar, directed graphs. In particular, it examines 27,000 urban street networks---represented as primal, nonplanar, weighted multidigraphs with possible self-loops---at multiple overlapping scales across the US, focusing on structure, connectedness, centrality, and resilience. It examines the street networks of every incorporated city and town, census urbanized area, and Zillow-defined neighborhood in the US. To do so, it uses OSMnx\endnote{OSMnx is freely available online at \url{https://github.com/gboeing/osmnx}}---a new street network research toolkit \citep{boeing_osmnx:_2017}---to download, model, and analyze these street networks at metropolitan, municipal, and neighborhood scales. These street networks and measures data sets have been compiled and shared in a public repository at the Harvard Dataverse\endnote{\label{note_data_repo}Data repository available online at \url{https://dataverse.harvard.edu/dataverse/osmnx-street-networks}} for other researchers to use.

The purpose of this paper is threefold. First, it describes and demonstrates a new methodology for easily and consistently acquiring, modeling, and analyzing large samples of street networks as nonplanar directed graphs. Second, it presents empirical findings of descriptive urban morphology for the street networks of every US city, urbanized area, and Zillow neighborhood. Third, it investigates with large sample sizes some previous smaller-sample findings in the research literature. This paper is organized as follows. In the next section, it discusses the data sources, tools, and methods used to collect, model, and analyze these street networks. Then it presents findings of the analyses at metropolitan, municipal, and neighborhood scales. Finally, it concludes with a discussion of these findings and their implications for street network analysis, urban morphology, and city planning.

\section{Methodology}

A network (also called a \emph{graph} in mathematics) comprises a set of nodes connected to one another by a set of edges. Street networks can be conceptualized as primal, directed, nonplanar graphs. A \emph{primal} street network represents intersections as nodes and street segments as edges. A \emph{directed} network has directed edges: that is, edge $uv$ points one-way from node $u$ to node $v$, but there need not exist a reciprocal edge $vu$. A \emph{planar} network can be represented in two dimensions with its edges intersecting only at nodes \citep{viana_simplicity_2013, fischer_spatial_2014}. Most street networks are nonplanar---due to grade-separated expressways, overpasses, bridges, tunnels, etc.---but most quantitative studies of urban street networks represent them as planar \citep[e.g.,][]{buhl_topological_2006, cardillo_structural_2006, barthelemy_modeling_2008, masucci_random_2009, strano_urban_2013} for tractability because bridges and tunnels are uncommon in some cities. Planar graphs may reasonably model the street networks of old European town centers, but poorly model the street networks of modern autocentric cities like Los Angeles or Shanghai with many grade-separated expressways, bridges, and underpasses \citep{boeing_planarity_2018}.

\subsection{Study Sites and Data Acquisition}

This study uses OSMnx to download, model, correct, analyze, and visualize street network graphs at metropolitan, municipal, and neighborhood scales. OSMnx is a Python-based research tool that easily downloads OpenStreetMap data for any place name, address, or polygon in the world, then constructs it into a spatially-embedded graph-theoretic object for analysis and visualization \citep{boeing_osmnx:_2017}. OpenStreetMap is a collaborative worldwide mapping project that makes its spatial data available via various APIs \citep{corcoran_analysing_2013, jokar_arsanjani_openstreetmap_2015}. These data are of high quality and compare favorably to CIA World Factbook estimates and US Census TIGER/Line data \citep{haklay_how_2010, over_generating_2010, zielstra_comparative_2011, maron_how_2015, wu_improving_2005, frizzelle_importance_2009}. In 2007, OpenStreetMap imported the TIGER/Line roads (2005 vintage) and since then, many community-led corrections and improvements have been made \citep{willis_openstreetmap_2008}. Many of these additions go beyond TIGER/Line's scope, including passageways between buildings, footpaths through parks, bike routes, and detailed feature attributes such as finer-grained street classifiers, speed limits, etc.

To define the study sites and their spatial boundaries, we use three sets of geometries. The first defines the metropolitan-scale study sites using the 2016 TIGER/Line shapefile of US census bureau urban areas. Each census-defined urban area comprises a set of tracts that meet a minimum density threshold \citep{u.s._census_bureau_2010_2010}. We retain only the \emph{urbanized areas} subset of these data (i.e., areas with greater than 50,000 population), discarding the small \emph{urban clusters} subset. The second set of geometries defines our municipal-scale study sites using 51 separate TIGER/Line shapefiles (again, 2016) of US census bureau \emph{places} within all 50 states plus DC. We discard the subset of \emph{census-designated places} (i.e., small unincorporated communities) in these data, while retaining every US city and town. The third set of geometries defines the neighborhood-scale study sites using 42 separate shapefiles from Zillow, a real estate database company. These shapefiles contain neighborhood boundaries for major cities in 41 states plus DC. This fairly new data set comprises nearly 7,000 neighborhoods, but as \citet{schernthanner_spatial_2016} point out, Zillow does not publish the methodology used to construct these boundaries. However, despite its newness it already has a track record in the academic literature: \citet{besbris_effect_2015} use Zillow boundaries to examine neighborhood stigma and \citet{albrecht_indicator_2014} use them to study neighborhood-level poverty in New York.

For each of these geometries, we use OSMnx to download the (drivable, public) street network within it, a process described in detail in \citet{boeing_osmnx:_2017} and summarized here. First OSMnx buffers each geometry by 0.5 km, then downloads the OpenStreetMap \enquote{nodes} and \enquote{ways} within this buffer. Next it constructs a street network graph from these data, corrects the topology, calculates street counts per node, then truncates the network to the original, desired polygon. OSMnx saves each of these networks to disk as GraphML and shapefiles. Finally, it calculates metric and topological measures for each network, summarized below. Such measures extend the toolkit commonly used in urban form studies \citep{talen_measuring_2003, ewing_travel_2010}.

\subsection{Street Network Measures}

Brief descriptions of these OSMnx-calculated measures are discussed here, but extended technical definitions and algorithms can be found in e.g. \citep{trudeau_introduction_1994, albert_statistical_2002, dorogovtsev_evolution_2002, brandes_network_2005, costa_characterization_2007, newman_structure_2003, newman_networks:_2010, barthelemy_spatial_2011, cranmer_navigating_2017}. The \emph{average street segment length} is a linear proxy for block size and specifies the network's grain. \emph{Node density} divides the node count by the network's area, while \emph{intersection density} excludes dead-ends to represent the density of street junctions. \emph{Edge density} divides the total directed network length by area, while \emph{street density} does the same for an undirected representation of the graph (to not double-count bidirectional streets). \emph{Average circuity} measures the ratio of edge lengths to the great-circle distances between the nodes these edges connect, indicating the street pattern's curvilinearity \citep[cf.][]{boeing_morphology_2018}.

The network's \emph{average node degree} quantifies connectedness in terms of the average number of edges incident to its nodes. The \emph{average streets per node} adapts this for physical form rather than directed circulation. It measures the average number of physical streets that emanate from each node (i.e., intersection or dead-end). The distribution and proportion of streets per node characterize the type, pervasiveness, and spatial dispersal of network connectedness and dead-ends. \emph{Connectivity} represents the fewest number of nodes or edges that will disconnect the network if they are removed and is thus an indicator of resilience. A network's \emph{average node connectivity} (ANC)---the mean number of internally node-disjoint paths between each pair of nodes---more usefully represents how many nodes must be removed on average to disconnect a randomly selected pair of nodes \citep{beineke_average_2002,dankelmann_bounds_2003}. Brittle points of vulnerability characterize networks with low average connectivity.

A node's \emph{clustering coefficient} represents the ratio between its neighbors' links and the maximum number of links that could exist between them \citep{jiang_topological_2004, opsahl_clustering_2009}. The weighted clustering coefficient weights this by edge length and the average clustering coefficient is the mean of the clustering coefficients of all the nodes. \emph{Betweenness centrality} evaluates how many of the network's shortest paths pass through some node (or edge) to indicate its importance \citep{barthelemy_betweenness_2004,huang_trajgraph:_2016, zhong_revealing_2017}. A network's \emph{maximum betweenness centrality} (MBC) measures the share of shortest paths that pass through the network's most important node: higher maximum betweenness centralities suggest networks more prone to inefficiency if this important choke point should fail. Finally, \emph{PageRank} ranks nodes based on the structure of incoming links and the rank of the source node \citep{brin_anatomy_1998, jiang_ranking_2009, agryzkov_algorithm_2012, chin_geographically_2015, gleich_pagerank_2015}.

In total, this study cross-sectionally analyzes 27,009 networks: 497 urbanized areas' street networks, 19,655 cities' and towns' street networks, and 6,857 neighborhoods' street networks. These sample sizes are larger than those of any previous similar study. The following section presents the findings of these analyses at metropolitan, municipal, and neighborhood scales.

\section{Results}

\subsection{Metropolitan-Scale Street Networks}

Table \ref{tab:measures_urban_areas} presents summary statistics for the entire data set of 497 urbanized areas. These urbanized areas span a wide range of sizes, from the Delano, CA Urbanized Area's 26 km\textsuperscript{2} to the New York--Newark, NY--NJ--CT Urbanized Area's 8,937 km\textsuperscript{2}. Thus, density and count-based measures demonstrate substantial variance. Further, these urbanized areas span a wide spectrum of terrains, development eras and paradigms, and cultures.

\begin{table*}[htbp]
	\centering
	\caption{Central tendency and statistical dispersion for selected measures of all US urbanized areas' street networks: $\mu$ is the mean, $\sigma$ is the standard deviation, and $D$ is the dispersion index $\frac{\sigma ^ 2}{\mu}$.}
	\label{tab:measures_urban_areas}
	\small
	\begin{tabular}{ l r r r r r r }
		\toprule
		measure                                          & $\mu$          & $\sigma$       & min            & median         & max            & $D$            \\
		\midrule
		Area (km\textsuperscript{2})                     & 460.657        & 858.125        & 25.685         & 184.898        & 8937.429       & 1598.539       \\
		Avg of the avg neighborhood degree               & 2.886          & 0.109          & 2.626          & 2.875          & 3.228          & 0.004          \\
		Avg of the avg weighted neighborhood degree      & 0.032          & 0.018          & 0.021          & 0.030          & 0.321          & 0.011          \\
		Avg circuity                                     & 1.076          & 0.019          & 1.023          & 1.074          & 1.140          & \textless0.001 \\
		Avg clustering coefficient                       & 0.042          & 0.009          & 0.015          & 0.042          & 0.071          & 0.002          \\
		Avg weighted clustering coefficient              & 0.002          & 0.001          & \textless0.001 & 0.001          & 0.006          & \textless0.001 \\
		Intersection count                               & 12582          & 26054          & 751            & 4593           & 307848         & 53949.814      \\
		Avg degree centrality                            & 0.001          & 0.001          & \textless0.001 & 0.001          & 0.007          & 0.001          \\
		Edge density (km/km\textsuperscript{2})          & 13.455         & 2.137          & 7.961          & 13.352         & 21.233         & 0.340          \\
		Avg edge length (m)                              & 158.588        & 17.653         & 117.341        & 157.332        & 223.080        & 1.965          \\
		Total edge length (km)                           & 6353           & 12625          & 427            & 2393           & 1.42e8         & 25089.459      \\
		Proportion of dead-ends                          & 0.213          & 0.055          & 0.077          & 0.207          & 0.416          & 0.014          \\
		Proportion of 3-way intersections                & 0.593          & 0.046          & 0.444          & 0.591          & 0.778          & 0.004          \\
		Proportion of 4-way intersections                & 0.187          & 0.063          & 0.054          & 0.178          & 0.422          & 0.021          \\
		Intersection density (per km\textsuperscript{2}) & 26.469         & 6.256          & 12.469         & 26.029         & 49.423         & 1.478          \\
		Average node degree                              & 5.153          & 0.302          & 4.307          & 5.143          & 6.056          & 0.018          \\
		$m$                                              & 40890          & 83678          & 2516           & 14955          & 981646         & 171238.406     \\
		$n$                                              & 16032          & 32585          & 874            & 5830           & 373309         & 66229.939      \\
		Node density (per km\textsuperscript{2})         & 33.628         & 7.641          & 17.675         & 33.071         & 61.655         & 1.736          \\
		Max PageRank value                               & 0.001          & 0.001          & \textless0.001 & 0.001          & 0.003          & \textless0.001 \\
		Min PageRank value                               & \textless0.001 & \textless0.001 & \textless0.001 & \textless0.001 & \textless0.001 & \textless0.001 \\
		Self-loop proportion                             & 0.008          & 0.008          & \textless0.001 & 0.006          & 0.071          & 0.008          \\
		Street density (km/km\textsuperscript{2})        & 7.262          & 1.221          & 4.217          & 7.171          & 11.797         & 0.205          \\
		Average street segment length (m)                & 161.331        & 17.765         & 119.573        & 160.288        & 225.920        & 1.956          \\
		Total street length (km)                         & 3480           & 7026           & 222            & 1269           & 79046          & 14185.880      \\
		Street segment count                             & 22011          & 45725          & 1281           & 7868           & 533757         & 94987.570      \\
		Average streets per node                         & 2.764          & 0.162          & 2.223          & 2.770          & 3.217          & 0.010          \\
		\bottomrule
	\end{tabular}
\end{table*}

\begin{figure*}[htbp]
	\centering
	\includegraphics[width=1\textwidth]{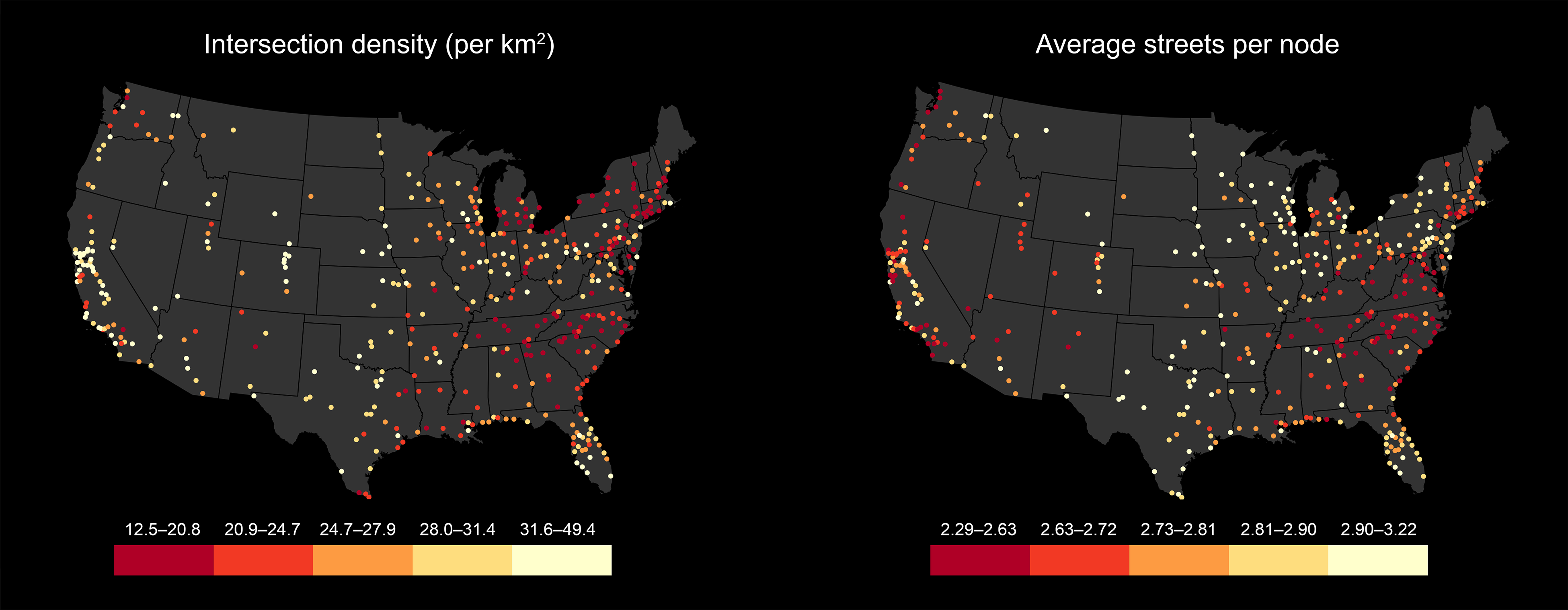}
	\caption{Intersection density and average streets per node per urbanized area in the contiguous US.}
	\label{fig:fig01}
\end{figure*}

\begin{figure*}[htbp]
	\centering
	\includegraphics[width=0.7\textwidth]{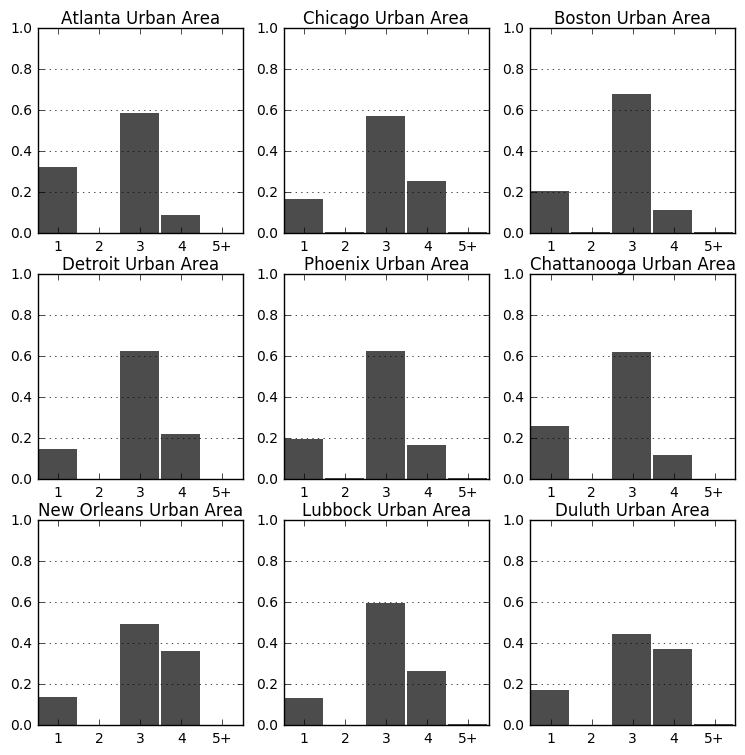}
	\caption{Distribution of node types in 9 urbanized areas, with number of streets emanating from the node on the $x$-axis and proportion of nodes of this type on the $y$-axis }
	\label{fig:fig02}
\end{figure*}

Nevertheless, looking across the data set provides a sense of the breadth of American metropolitan street networks. New York's urbanized area---America's largest---has 373,309 intersections and 79 million meters of linear street (or 417,570 and 83.4 million if including service roads). Delano, CA's urbanized area---America's smallest---has 874 intersections and 222,328 meters of linear street (or 964 and 231,000 meters if including service roads). The typical American urbanized area is approximately 185 km\textsuperscript{2} in land area, has 5,830 intersections, and 1.3 million linear meters of street. Its street network is about 7.4\% more circuitous than straight-line as-the-crow-flies edges between nodes would be. The most circuitous network is 14\% more circuitous than straight-line would be, and the least is only 2\%. Looking at density, grain, and connectivity in the typical urbanized area, the average street segment length (a proxy for block size) is 160 meters. The longest average street segment is the 226-meter average of urbanized Danbury, CT. Puerto Rican cities hold the top four positions for shortest average street segment length, but among the 50 states plus DC, the shortest average street segment is the 125.3-meter average of urbanized Tracy, CA, indicating a fine-grained network. The urbanized area of Portland, Oregon, with its famously compact walkable blocks, ranks second at 125.5 meters.

The typical urbanized area has 26 intersections per km\textsuperscript{2}. Both the densest and the sparsest are in the Deep South: the sparsest has 12.5 (Gainesville, GA urbanized area) and the densest has 49.4 (New Orleans urbanized area). However, New Orleans is an anomaly in the Deep South. Figure \ref{fig:fig01} depicts the intersection density of each American urbanized area: the highest intersection densities concentrate west of the Mississippi River, while the lowest concentrate in a belt running from Louisiana, through the Carolinas and Appalachians, and into New England. In general, only the largest cities on the east coast (e.g., Boston, New York, Philadelphia, Washington) and Florida escape this trend.

The distribution of node types (i.e., intersections and dead ends) provides an indicator of network connectedness. The typical urbanized area averages 2.8 streets per intersection: many 3-way intersections, fewer dead-ends, and even fewer 4-way intersections. The gridlike San Angelo, TX urbanized area has the most streets per node (3.2) on average, and (outside of Puerto Rico, which contains the seven lowest urbanized areas) the sprawling, disconnected Lexington Park, MD urbanized area has the fewest (2.2). These fit the trend seen in the spatial distribution across the US in Figure \ref{fig:fig01}: urbanized areas in the Great Plains and Midwest have particularly high numbers of streets per node on average, indicating more gridlike, connected networks. Cities in the southern and western US tend to have fewer streets per node, reflecting more dead-ends and a disconnected network. This finding is discussed in more detail in the upcoming section.

In the typical urbanized area, 18\% of nodes are 4-way intersections, 59\% are 3-way intersections, and 21\% are dead-ends. However, this distribution varies somewhat: examining a small sample of nine urbanized areas, chosen to maximize variance, reveals this in clearer detail. In Figure \ref{fig:fig02}, urban Atlanta and Chattanooga have very high proportions of dead-ends---each over 30\% of all nodes---and very few 4-way intersections, indicating a disconnected street pattern. The urbanized areas of Phoenix, Boston, Detroit, and Chattanooga have particularly high proportions of 3-way intersections, each over 60\%, indicating a prevalence of \emph{T}-intersections. Conversely, Chicago, New Orleans, Duluth, and Lubbock have high proportions of 4-way intersections, indicating more gridlike connected networks. But what is perhaps most notable about Figure \ref{fig:fig02} is that these nine urbanized areas, despite being chosen to maximize variance, are overwhelmingly similar to each other. Every large American urban agglomeration is characterized by a preponderance of 3-way intersections.

The relationship between fine-grained networks and connectedness/gridness is not, however, clear-cut: intersection density has only a weak, positive linear relationship with the proportion of 4-way intersections in the urbanized area ($r^{2}=0.17$). But the relationship between network circuity and gridness is somewhat clearer: average circuity has a negative linear relationship with the proportion of 4-way intersections ($r^{2}=0.43$).

The dispersion index $D$ in Table \ref{tab:measures_urban_areas} demonstrates the heterogeneity of each indicator across the data set. Several \enquote{families} of indicators can be discerned by their heterogeneity. For instance, counts and totals such as $m$, $n$, and the total street length are extremely heterogeneous. Densities and average distances such as intersection density and the average street segment length exhibit only moderate heterogeneity. Finally, several topological measures such as the average clustering coefficient and PageRank are extremely homogeneous. Due to the substantial variation in urbanized area size, from 25 to 9,000 km\textsuperscript{2}, the preceding analysis covers a wide swath of metropolitan types. To better compare apples-to-apples, Table \ref{tab:largest_urbanized_areas} focuses on the 30 largest urbanized areas cross-sectionally to examine their metric and topological measures. This provides more consistent spatial scales and extents, while offering a window into the similarities and differences in the built forms of America's largest agglomerations.

Among these urbanized areas, Milwaukee has the least circuitous network (6\% more circuitous than straight-line edges would be), and Orlando has the most (12\%). San Juan and Atlanta have the fewest streets per node on average (2.36 and 2.45, respectively), while Milwaukee has the most (3.03). Cincinnati has both the lowest intersection density (18/km\textsuperscript{2}) and street density (6.1 km/km\textsuperscript{2}) while Denver has the highest intersection density (40.6/km\textsuperscript{2}) and Miami and Los Angeles have the highest street density (10.6 km/km\textsuperscript{2}, apiece). In other words, Cincinnati has a particularly coarse-grained network with few connections and paths. The average street segment length, a proxy for block size, also reflects this: Cincinnati has the second highest (186 m), bested only by Cleveland (198 m). In contrast, the two lowest are Denver's 138-meter average and San Juan's 131-meter average.

These metropolitan analyses consider trends in the built form at the scale of broad human systems and urbanized regions. However, they aggregate multiple heterogeneous municipalities and neighborhoods---the scales of human life, urban design projects, and planning jurisdictions---into single units of analysis. To disaggregate and analyze finer characteristics, the following sections examine municipal- and neighborhood-scale street networks.

\end{multicols}
\begin{sidewaystable*}[p]
\centering
\caption{Selected measures of the 30 largest (by land area) urbanized areas' street networks.}
\label{tab:largest_urbanized_areas}
\small
\begin{tabular}{p{2.5cm} p{1.5cm} p{1.5cm} p{1.5cm} p{1.55cm} p{1.5cm} p{1.5cm} p{1.5cm} p{1.5cm} p{1.5cm} p{1.5cm}}
	\hline
	urban area core city & land area km\textsuperscript{2} & avg circuity & avg clustering coefficient & dead-end ratio & 3-way ratio & 4-way ratio & intersect density/ km\textsuperscript{2} & street density km/km\textsuperscript{2} & avg street length (m) & avg streets/ node \\
	\hline
	New York             & 8937          & 1.06         & 0.04                       & 0.18           & 0.62        & 0.20        & 34.44                 & 8.84                  & 148                   & 2.86             \\
	Atlanta              & 6850          & 1.10         & 0.04                       & 0.32           & 0.58        & 0.09        & 18.39                 & 6.16                  & 186                   & 2.45             \\
	Chicago              & 6325          & 1.07         & 0.04                       & 0.17           & 0.57        & 0.25        & 27.05                 & 7.77                  & 163                   & 2.92             \\
	Philadelphia         & 5132          & 1.08         & 0.05                       & 0.17           & 0.63        & 0.20        & 26.65                 & 7.30                  & 159                   & 2.87             \\
	Boston               & 4852          & 1.09         & 0.05                       & 0.20           & 0.68        & 0.11        & 24.23                 & 6.44                  & 154                   & 2.71             \\
	Dallas               & 4612          & 1.07         & 0.05                       & 0.15           & 0.61        & 0.23        & 34.16                 & 9.16                  & 156                   & 2.95             \\
	Los Angeles          & 4497          & 1.06         & 0.03                       & 0.21           & 0.56        & 0.22        & 39.45                 & 10.59                 & 151                   & 2.82             \\
	Houston              & 4303          & 1.08         & 0.04                       & 0.20           & 0.57        & 0.22        & 33.49                 & 8.62                  & 145                   & 2.83             \\
	Detroit              & 3461          & 1.07         & 0.04                       & 0.15           & 0.63        & 0.22        & 31.10                 & 8.56                  & 159                   & 2.93             \\
	Washington           & 3424          & 1.09         & 0.04                       & 0.26           & 0.56        & 0.17        & 31.22                 & 8.26                  & 146                   & 2.66             \\
	Miami                & 3204          & 1.10         & 0.05                       & 0.17           & 0.59        & 0.23        & 40.54                 & 10.61                 & 149                   & 2.89             \\
	Phoenix              & 2968          & 1.09         & 0.05                       & 0.20           & 0.62        & 0.17        & 35.31                 & 9.10                  & 150                   & 2.77             \\
	Minneapolis          & 2647          & 1.08         & 0.05                       & 0.19           & 0.57        & 0.23        & 29.54                 & 8.68                  & 167                   & 2.84             \\
	Seattle              & 2617          & 1.07         & 0.03                       & 0.30           & 0.54        & 0.16        & 31.57                 & 8.20                  & 143                   & 2.57             \\
	Tampa                & 2479          & 1.10         & 0.05                       & 0.20           & 0.58        & 0.21        & 31.35                 & 8.46                  & 153                   & 2.83             \\
	St. Louis            & 2392          & 1.10         & 0.04                       & 0.22           & 0.62        & 0.15        & 29.68                 & 8.16                  & 154                   & 2.73             \\
	Pittsburgh           & 2345          & 1.09         & 0.04                       & 0.23           & 0.60        & 0.16        & 23.57                 & 6.71                  & 165                   & 2.72             \\
	San Juan             & 2245          & 1.11         & 0.02                       & 0.36           & 0.56        & 0.08        & 26.57                 & 6.43                  & 131                   & 2.36             \\
	Cincinnati           & 2040          & 1.07         & 0.03                       & 0.31           & 0.54        & 0.14        & 17.96                 & 6.10                  & 186                   & 2.51             \\
	Cleveland            & 2004          & 1.07         & 0.04                       & 0.19           & 0.66        & 0.14        & 19.13                 & 6.51                  & 198                   & 2.76             \\
	Charlotte            & 1920          & 1.08         & 0.04                       & 0.30           & 0.57        & 0.11        & 21.00                 & 6.43                  & 170                   & 2.51             \\
	San Diego            & 1897          & 1.08         & 0.03                       & 0.28           & 0.54        & 0.17        & 28.89                 & 8.32                  & 159                   & 2.62             \\
	Baltimore            & 1857          & 1.09         & 0.04                       & 0.23           & 0.59        & 0.17        & 27.72                 & 7.56                  & 152                   & 2.72             \\
	Indianapolis         & 1828          & 1.08         & 0.05                       & 0.23           & 0.59        & 0.17        & 27.62                 & 7.63                  & 157                   & 2.70             \\
	Kansas City          & 1756          & 1.06         & 0.04                       & 0.21           & 0.58        & 0.20        & 32.09                 & 8.57                  & 152                   & 2.79             \\
	Denver               & 1729          & 1.07         & 0.05                       & 0.20           & 0.57        & 0.22        & 40.60                 & 9.84                  & 138                   & 2.84             \\
	Orlando              & 1548          & 1.11         & 0.06                       & 0.20           & 0.61        & 0.18        & 26.30                 & 7.44                  & 163                   & 2.79             \\
	San Antonio          & 1547          & 1.07         & 0.05                       & 0.17           & 0.60        & 0.21        & 28.33                 & 7.91                  & 162                   & 2.87             \\
	Nashville            & 1460          & 1.08         & 0.03                       & 0.27           & 0.59        & 0.14        & 19.08                 & 6.10                  & 181                   & 2.60             \\
	Milwaukee            & 1413          & 1.06         & 0.06                       & 0.14           & 0.55        & 0.30        & 28.27                 & 7.81                  & 157                   & 3.03 \\
	\bottomrule
\end{tabular}
\end{sidewaystable*}

\begin{multicols}{2}

\subsection{Municipal-Scale Street Networks}

\begin{figure*}[htbp]
	\centering
	\includegraphics[width=1\textwidth]{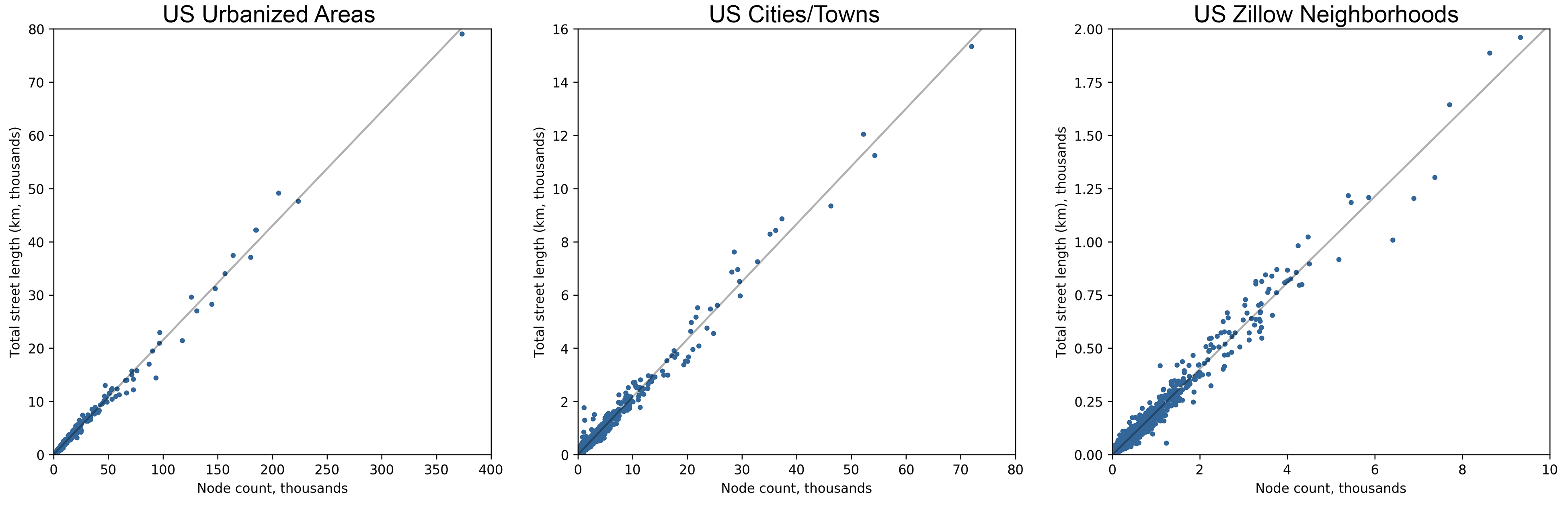}
	\caption{The linear relationship between total street length and number of nodes in the street networks of every US urbanized area, city/town, and Zillow neighborhood.}
	\label{fig:fig03}
\end{figure*}

Table \ref{tab:measures_cities} presents summary statistics of street network characteristics across the entire data set of 19,655 cities and towns---every incorporated city and town in the US. Following recent work by \citet{barthelemy_modeling_2008} and \citet{strano_urban_2013}, we examine the relationship between the total street length $L$ and the number of nodes $n$ across different cities. The former proposed a model of city network evolution in which $L$ and $n$ scale nonlinearly as $n^{1/2}$, and the latter suggested that this relationship applies cross-sectionally, using an empirical sample of ten European cities. However, the latter's small sample size may limit the generalizability of this finding. We examine the relationship between $L$ and $n$ across every US city and town and instead find a strong linear relationship ($r^{2}=0.98$), as depicted in Figure \ref{fig:fig03}. We also find a similar linear relationship at the metropolitan ($r^{2}=0.99$) and neighborhood ($r^{2}=0.98$) scales.

\begin{table*}
	\centering
	\caption{Central tendency and statistical dispersion for selected measures of all incorporated cities and towns in the US: $\mu$ is the mean, $\sigma$ is the standard deviation, and $D$ is the dispersion index $\frac{\sigma ^ 2}{\mu}$.}
	\label{tab:measures_cities}
	\small
	\begin{tabular}{ l r r r r r r }
		\toprule
		measure                                          & $\mu$   & $\sigma$& min            & median         & max       & $D$      \\
		\midrule
		Area (km\textsuperscript{2})                     & 16.703  & 107.499 & 0.039          & 3.918          & 7434.258  & 691.860  \\
		Avg of the avg neighborhood degree               & 2.940   & 0.297   & 0.400          & 2.953          & 3.735     & 0.030    \\
		Avg of the avg weighted neighborhood degree      & 0.033   & 0.141   & \textless0.001 & 0.029          & 9.357     & 0.607    \\
		Avg circuity                                     & 1.067   & 0.159   & 1.000          & 1.055          & 20.452    & 0.024    \\
		Avg clustering coefficient                       & 0.048   & 0.041   & \textless0.001 & 0.04           & 1.000     & 0.035    \\
		Avg weighted clustering coefficient              & 0.010   & 0.018   & \textless0.001 & 0.005          & 0.524     & 0.033    \\
		Intersection count                               & 324     & 1266    & 0              & 83             & 62996     & 4951.293 \\
		Avg degree centrality                            & 0.093   & 0.136   & \textless0.001 & 0.052          & 2.667     & 0.199    \\
		Edge density (km/km\textsuperscript{2})          & 12.654  & 6.705   & 0.006          & 11.814         & 58.603    & 3.553    \\
		Avg edge length (m)                              & 161.184 & 80.769  & 25.822         & 144.447        & 3036.957  & 40.473   \\
		Total edge length (km)                           & 159.067 & 578.521 & 0.052          & 40.986         & 24728.326 & 2104.061 \\
		Proportion of dead-ends                          & 0.192   & 0.093   & \textless0.001 & 0.184          & 1.000     & 0.045    \\
		Proportion of 3-way intersections                & 0.572   & 0.11    & \textless0.001 & 0.579          & 1.000     & 0.021    \\
		Proportion of 4-way intersections                & 0.237   & 0.129   & \textless0.001 & 0.217          & 1.000     & 0.070    \\
		Intersection density (per km\textsuperscript{2}) & 29.363  & 21.607  & \textless0.001 & 24.719         & 259.647   & 15.900   \\
		Average node degree                              & 5.251   & 0.668   & 0.800          & 5.268          & 7.166     & 0.085    \\
		$m$                                              & 1046    & 3924    & 2              & 275            & 176161    & 14714.556\\
		$n$                                              & 401     & 1516    & 2              & 103            & 71993     & 5734.363 \\
		Node density (per km\textsuperscript{2})         & 35.449  & 24.409  & 0.047          & 30.718         & 296.740   & 16.807   \\
		Max PageRank value                               & 0.034   & 0.046   & \textless0.001 & 0.021          & 0.870     & 0.062    \\
		Min PageRank value                               & 0.005   & 0.018   & \textless0.001 & 0.002          & 0.500     & 0.060    \\
		Self-loop proportion                             & 0.005   & 0.015   & \textless0.001 & \textless0.001 & 1.000     & 0.042    \\
		Street density (km/km\textsuperscript{2})        & 6.528   & 3.435   & 0.003          & 6.109          & 29.302    & 1.807    \\
		Average street segment length (m)                & 162.408 & 81.035  & 25.822         & 145.479        & 3036.957  & 40.433   \\
		Total street length (km)                         & 86.096  & 331.048 & 0.026          & 21.005         & 15348.008 & 1272.917 \\
		Street segment count                             & 558     & 2208    & 1.000          & 140            & 107393    & 8745.983 \\
		Average streets per node                         & 2.851   & 0.282   & 1.000          & 2.852          & 4.000     & 0.028    \\
		\bottomrule
	\end{tabular}
\end{table*}

Previous findings \citep[e.g.,][]{masucci_random_2009,gudmundsson_entropy_2013} suggest street segment lengths in an urban network follow a power-law distribution. We find that these networks instead generally follow lognormal-style right-skewed distributions. This makes theoretical sense as most street networks are not truly scale-free: for example, a typical street network might comprise very few very long street segments (e.g., 1 km), more medium-length segments (e.g., 250 m), many short segments (e.g., 80 m), but very few very short segments (e.g., 10 m). To test this, we fit a set of candidate distributions to the street segment lengths of each city/town. These distributions comprise the lognormal, Gumbel, gamma, exponentiated Weibull, Fréchet, power-law, uniform, and exponential distributions. We then assess these fits via the Akaike information criterion to compare their relative performance in modeling the observed data. Power-law distributions provide the best fit for only 3\% of these cities. In contrast, the exponentiated Weibull distribution provides the best fit 52\% of the time, followed by the Gumbel (21\%), gamma (10\%), and lognormal (7\%) distributions.

An exception to this general pattern, of course, lies in consistently-sized orthogonal grids filling a city's incorporated spatial extents. Such distributions are extremely peaked around a single value: the linear length of a grid block. We find that such cities are not uncommon in the US, particularly between the Mississippi River and the Rocky Mountains: the Great Plains states are characterized by a unique street network form that is both orthogonal and reasonably dense. The former is partly the result of topography (flat terrain that allows idealized grids) and design history (rapid platting and development during the late nineteenth century) that favor orthogonal grids, as discussed earlier. The latter results from the fact that most towns across the Great Plains exhibit minimal suburban sprawl. Thus, municipal boundaries snugly embrace the gridlike street network, without extending to accommodate a vast peripheral belt of twentieth century sprawl, circuity, and \enquote{loops and lollipops} \citep{southworth_streets_1997} that characterizes cities in e.g. California that were settled in the same era but later subjected to substantial suburbanization.

\begin{figure*}[htbp]
	\centering
	\includegraphics[width=1\textwidth]{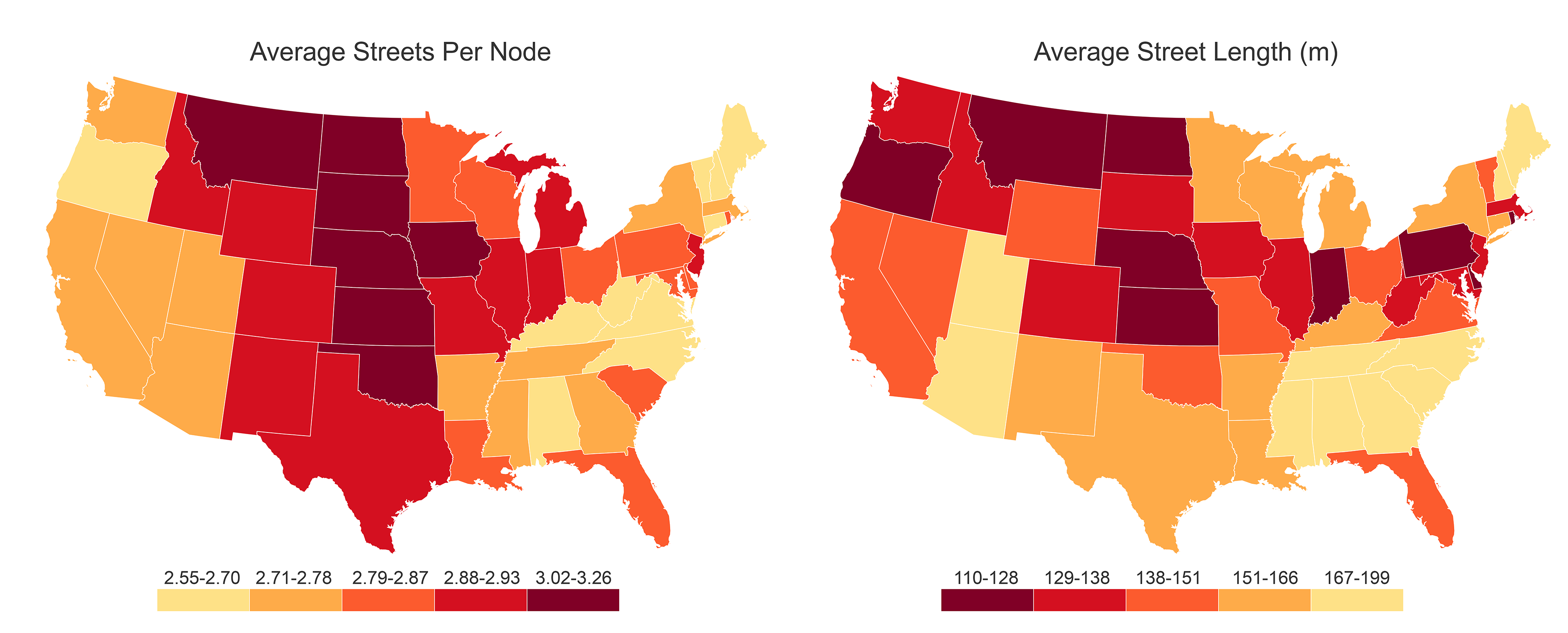}
	\caption{Contiguous US states by median of mean streets per node and by median of mean street segment length in municipal street networks.}
	\label{fig:fig04}
\end{figure*}

\begin{table*}[htbp]
	\centering
	\caption{Median values, aggregated by state plus DC, of selected measures of the municipal-scale street networks for every city and town in the US.}
	\label{tab:medians_state_aggregate}
	\footnotesize
	\begin{tabular}{ l r r r r }
		\toprule
		State & Intersection density (per km\textsuperscript{2}) & Avg streets per node & Avg circuity & Avg street segment length \\
		\midrule
		AK    &  1.28                       & 2.43                 & 1.10         & 223.50                    \\
		AL    &  9.70                       & 2.64                 & 1.07         & 190.81                    \\
		AR    & 15.75                       & 2.78                 & 1.06         & 166.32                    \\
		AZ    & 12.45                       & 2.77                 & 1.08         & 171.80                    \\
		CA    & 32.58                       & 2.74                 & 1.07         & 143.79                    \\
		CO    & 29.26                       & 2.88                 & 1.06         & 136.68                    \\
		CT    & 28.05                       & 2.70                 & 1.07         & 165.87                    \\
		DC    & 58.91                       & 3.26                 & 1.04         & 122.23                    \\
		DE    & 25.30                       & 2.80                 & 1.06         & 127.80                    \\
		FL    & 26.26                       & 2.87                 & 1.07         & 150.75                    \\
		GA    & 15.25                       & 2.78                 & 1.07         & 177.50                    \\
		HI    &  8.00                       & 2.42                 & 1.07         & 177.93                    \\
		IA    & 24.08                       & 3.02                 & 1.04         & 129.36                    \\
		ID    & 33.85                       & 2.91                 & 1.06         & 132.08                    \\
		IL    & 29.02                       & 2.93                 & 1.05         & 137.77                    \\
		IN    & 35.25                       & 2.93                 & 1.05         & 125.72                    \\
		KS    & 43.94                       & 3.14                 & 1.04         & 124.39                    \\
		KY    & 25.12                       & 2.68                 & 1.07         & 151.28                    \\
		LA    & 17.14                       & 2.79                 & 1.06         & 162.62                    \\
		MA    & 32.33                       & 2.76                 & 1.07         & 135.98                    \\
		MD    & 28.67                       & 2.79                 & 1.07         & 133.69                    \\
		ME    &  7.69                       & 2.67                 & 1.07         & 198.93                    \\
		MI    & 20.93                       & 2.90                 & 1.05         & 153.50                    \\
		MN    & 18.96                       & 2.87                 & 1.06         & 152.92                    \\
		MO    & 29.87                       & 2.89                 & 1.06         & 138.29                    \\
		MS    & 14.76                       & 2.75                 & 1.06         & 174.86                    \\
		MT    & 38.94                       & 3.11                 & 1.04         & 126.89                    \\
		NC    & 19.28                       & 2.65                 & 1.06         & 166.69                    \\
		ND    & 34.28                       & 3.07                 & 1.04         & 123.93                    \\
		NE    & 45.89                       & 3.16                 & 1.04         & 119.79                    \\
		NH    & 12.22                       & 2.69                 & 1.10         & 175.88                    \\
		NJ    & 44.98                       & 2.88                 & 1.04         & 130.79                    \\
		NM    & 18.50                       & 2.93                 & 1.05         & 152.02                    \\
		NV    & 13.86                       & 2.77                 & 1.07         & 147.35                    \\
		NY    & 21.89                       & 2.75                 & 1.06         & 156.88                    \\
		OH    & 25.23                       & 2.80                 & 1.05         & 142.08                    \\
		OK    & 28.22                       & 3.03                 & 1.05         & 139.50                    \\
		OR    & 35.08                       & 2.69                 & 1.06         & 121.18                    \\
		PA    & 35.69                       & 2.87                 & 1.05         & 128.34                    \\
		RI    & 56.23                       & 2.86                 & 1.05         & 110.35                    \\
		SC    & 18.76                       & 2.81                 & 1.06         & 169.21                    \\
		SD    & 32.01                       & 3.12                 & 1.04         & 130.75                    \\
		TN    & 13.62                       & 2.71                 & 1.07         & 192.83                    \\
		TX    & 23.85                       & 2.92                 & 1.05         & 160.44                    \\
		UT    & 12.58                       & 2.71                 & 1.06         & 191.04                    \\
		VA    & 25.18                       & 2.63                 & 1.08         & 145.65                    \\
		VT    & 18.91                       & 2.55                 & 1.08         & 145.18                    \\
		WA    & 28.71                       & 2.75                 & 1.06         & 134.02                    \\
		WI    & 17.87                       & 2.81                 & 1.06         & 156.19                    \\
		WV    & 28.45                       & 2.67                 & 1.08         & 136.57                    \\
		WY    & 23.48                       & 2.92                 & 1.06         & 143.63                    \\
		\bottomrule
	\end{tabular}
\end{table*}

For example, if we measure connectedness in terms of the average number of streets per node at the city-scale and then aggregate these cities by state (Table \ref{tab:medians_state_aggregate}), we find Nebraska, Kansas, South Dakota, Montana, North Dakota, Oklahoma, and Iowa have, in order, the highest medians (Figure \ref{fig:fig04}). This indicates the most gridlike networks. If we measure intersection density at the city-scale and then aggregate these cities by state, we find Rhode Island, Nebraska, New Jersey, Kansas, and Montana have, in order, the highest medians. We again see three Great Plains states near the top alongside small, densely populated East Coast states. Nebraska also has the smallest block sizes (measured via the proxy of average street segment length) while the largest concentrate in the Deep South, upper New England, and Utah (Figure \ref{fig:fig04}).

However, municipal boundaries vary greatly in their extents around the built-up area. While Rhode Island averages 56 intersections/km\textsuperscript{2} in its cities and towns, Alaska averages only 1.3, because the latter's municipal boundaries often extend thousands of km\textsuperscript{2} beyond the actual built-up area. In fact, Alaska has four cities (Anchorage, Juneau, Sitka, and Wrangell) with such large municipal extents that their land areas exceed that of the state of Rhode Island. These state-level aggregations of municipal street network characteristics show clear variation across the country that reflect topography, economies, culture, planning paradigms, and settlement eras. But they also aggregate and thus obfuscate the variation within each state and within each city. To explore these smaller-scale differences, the following section examines street networks at the neighborhood scale.

\subsection{Neighborhood-Scale Street Networks}

We have thus far examined every urban street network in the US at the metropolitan and municipal scales. While the metropolitan scale captures the emergent character of the wider region's complex system, and the municipal scale captures planning decisions made by a single city government, the neighborhood best represents the scale of individual urban design interventions into the urban form. Further, this scale more commonly reflects individual designs, eras, and paradigms in street network development than the \enquote{many hands, many eras} evolution of form at larger scales.

\begin{table*}[htbp]
	\centering
	\caption{Central tendency and statistical dispersion for selected measures of all the neighborhood-scale street networks: $\mu$ is the mean, $\sigma$ is the standard deviation, and $D$ is the dispersion index $\frac{\sigma ^ 2}{\mu}$.}
	\label{tab:measures_neighborhoods}
	\small
	\begin{tabular}{ l r r r r r r }
		\toprule
		measure                                          & $\mu$   & $\sigma$& min            & median         & max      & $D$     \\
		\midrule
		Area (km\textsuperscript{2})                     & 5.322   & 15.463  & 0.008          & 1.738          & 323.306  & 44.928  \\
		Avg of the avg neighborhood degree               & 2.598   & 0.436   & \textless0.001 & 2.670          & 3.632    & 0.073   \\
		Avg of the avg weighted neighborhood degree      & 0.031   & 0.041   & \textless0.001 & 0.029          & 2.991    & 0.054   \\
		Avg circuity                                     & 1.080   & 0.411   & 1.000          & 1.044          & 24.290   & 0.157   \\
		Avg clustering coefficient                       & 0.044   & 0.055   & \textless0.001 & 0.034          & 1.000    & 0.069   \\
		Avg weighted clustering coefficient              & 0.010   & 0.027   & \textless0.001 & 0.005          & 0.799    & 0.076   \\
		Intersection count                               & 173     & 379     & 0              & 76             & 8371     & 829.528 \\
		Avg degree centrality                            & 0.130   & 0.270   & 0.001          & 0.054          & 4.000    & 0.561   \\
		Edge density (km/km\textsuperscript{2})          & 17.569  & 7.095   & 0.025          & 18.152         & 59.939   & 2.866   \\
		Avg edge length (m)                              & 142.279 & 59.182  & 8.447          & 133.848        & 2231.331 & 24.617  \\
		Total edge length (km)                           & 71.369  & 166.566 & 0.017          & 29.880         & 3563.409 & 388.743 \\
		Proportion of dead-ends                          & 0.170   & 0.131   & \textless0.001 & 0.145          & 1.000    & 0.101   \\
		Proportion of 3-way intersections                & 0.559   & 0.146   & \textless0.001 & 0.574          & 1.000    & 0.038   \\
		Proportion of 4-way intersections                & 0.275   & 0.176   & \textless0.001 & 0.234          & 1.000    & 0.112   \\
		Intersection density (per km\textsuperscript{2}) & 49.497  & 28.330  & \textless0.001 & 46.430         & 444.355  & 16.216  \\
		Average node degree                              & 4.675   & 0.836   & 0.545          & 4.736          & 7.283    & 0.150   \\
		$m$                                              & 5201    & 1185    & 1              & 217            & 27289    & 2694.171\\
		$n$                                              & 208     & 459     & 2              & 90             & 9327     & 1014.643\\
		Node density (per km\textsuperscript{2})         & 58.677  & 31.802  & 0.063          & 55.626         & 499.900  & 17.237  \\
		Max PageRank value                               & 0.055   & 0.086   & \textless0.001 & 0.026          & 0.889    & 0.133   \\
		Min PageRank value                               & 0.010   & 0.041   & \textless0.001 & 0.002          & 0.500    & 0.161   \\
		Self-loop proportion                             & 0.007   & 0.034   & \textless0.001 & \textless0.001 & 1.000    & 0.177   \\
		Street density (km/km\textsuperscript{2})        & 9.744   & 4.085   & 0.013          & 9.882          & 33.737   & 1.712   \\
		Average street segment length (m)                & 143.664 & 60.023  & 7.376          & 134.877        & 2231.331 & 25.078  \\
		Total street length (km)                         & 40.049  & 93.987  & 0.009          & 16.248         & 1960.643 & 220.569 \\
		Street segment count                             & 288     & 656     & 1              & 119            & 14754    & 1491.595\\
		Average streets per node                         & 2.925   & 0.408   & 1.000          & 2.944          & 4.026    & 0.057   \\
		\bottomrule
	\end{tabular}
\end{table*}

Table \ref{tab:measures_neighborhoods} presents summary statistics for these 6,857 neighborhoods. Compared to the metropolitan and municipal scales, we see much greater variance here, as expected, given the smaller network sizes at the neighborhood scale. A few neighborhoods have no intersections within their Zillow-defined boundaries, resulting in a minimum intersection density of 0 across the data set. Meanwhile, the small neighborhood of Cottages North in Davis, California has the highest intersection density in the country, 444/km\textsuperscript{2}, largely an artifact of its small area as the denominator. Nationwide, the typical neighborhood averages 2.9 streets per intersection, reflecting the prevalence of 3-way intersections in the US, discussed earlier. The median proportions of each node type are 14.5\% for dead-ends, 57.4\% for 3-way intersections, and 23.4\% for 4-way intersections. The typical neighborhood averages 135-meter street segment lengths and 46.4 intersections per km\textsuperscript{2}.

\begin{figure*}[htbp]
	\centering
	\includegraphics[width=0.8\textwidth]{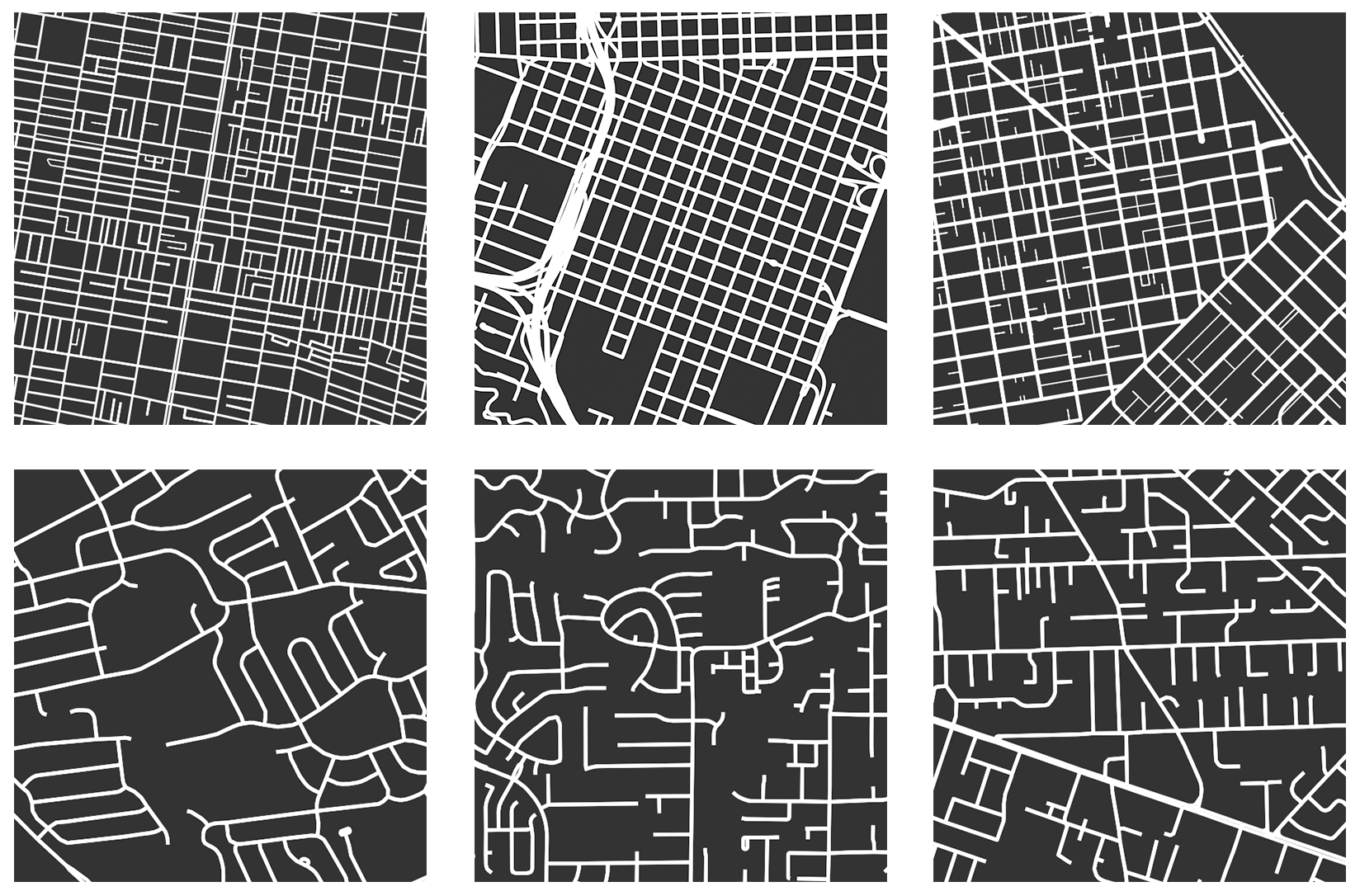}
	\caption{Square-mile comparisons of central cities and their suburbs. Left: top, downtown Philadelphia, PA; bottom, its suburb, King of Prussia. Middle: top, downtown Portland, OR; bottom, its suburb, Beaverton. Right: top, downtown San Francisco, CA; bottom, its suburb, Concord.}
	\label{fig:fig05}
\end{figure*}

Due to the extreme values seen---resulting from the large variance in neighborhood size---we can filter the data set to examine only large neighborhoods (i.e., with area greater than the median value across the data set). In this filtered set, the five neighborhoods with the highest intersection densities are all in central Philadelphia. Central neighborhoods are common at the top of this list, including Point Breeze, Philadelphia; Central Boston; Central City, New Orleans; Downtown Tampa; and Downtown Portland. The three neighborhoods with the lowest intersection densities are on the outskirts of Anchorage, Alaska. In the filtered set, the greatest average numbers of streets per node tend to be in older neighborhoods with orthogonal grids, such as Virginia Park, Tampa; Outer Sunset, San Francisco; and New Orleans' French Quarter. The neighborhoods with the lowest tend to be sprawling and often hilly suburbs far from the urban core, such as Scholl Canyon in Glendale, CA or Sonoma Ranch in San Antonio, TX.

To illustrate these morphologies, Figure \ref{fig:fig05} compares one square mile of the centers of Philadelphia, Portland, and San Francisco to one square mile of each of their suburbs. The connectedness and fine grain of the central cities are clear, as are the disconnectivity and coarse grain of their suburbs. In fact, the suburbs have more in common with one another---despite being hundreds or thousands of miles apart---than they do with their central city neighbors, suggesting that land use and an era's prevailing design paradigm is paramount to geographical localism and regional context. The top row of Figure \ref{fig:fig05} represents an era of planning and development that preceded the automobile, while the bottom row reflects the exclusionary zoning and mid-to-late twentieth century era of automobility in residential suburb design---namely the \enquote{loops and lollipops} and \enquote{lollipops on a stick} design patterns \citep{southworth_streets_1997}.

Finally, we briefly take a closer look at San Francisco, CA's neighborhoods alone for a clear cross-sectional analysis with consistent geography to examine resilience through the MBC and ANC measures. Due to its highly connected orthogonal grid, the Outer Sunset neighborhood has the lowest MBC---only 9.6\% of all shortest paths pass through its most important node. By contrast, 36\% of Chinatown's shortest paths pass through its most important node, and in Twin Peaks it is 37\%. In Chinatown, this is the result of a small neighborhood comprising only a few streets and that these streets are one-way, forcing paths through few routing options. In Twin Peaks, this is the result of hilly terrain and a disconnected network forcing paths through a small set of chokepoints that link separate subsections of the network. If a large number of shortest paths rely on a single node, the network is more prone to failure or inefficiency given a single point of failure.

In San Francisco, Twin Peaks' network has the lowest ANC: on average only 1.05 nodes must be removed to disconnect a randomly selected pair of nodes. Outer Sunset has the highest ANC, 3.2. These findings conform to the above descriptions of these networks. However, some central San Francisco orthogonal grid networks with many 4-way intersections---such as Downtown, Chinatown, and the Financial District---have surprisingly low ANCs: 1.5, 1.3, and 1.6 respectively. These neighborhoods comprise primarily one-way streets. Although they have dense, highly connected networks, they can be easily disconnected given that (automobile) traffic cannot flow bidirectionally. These three neighborhoods also exhibit the greatest increase in ANC if all their edges are made undirected: Chinatown's increases 87\%, Downtown's 80\%, and the Financial District's 75\%. By contrast, Outer Sunset's street network sees only a 6\% increase due to it already comprising primarily bidirectional streets. Targeted conversion of one-way streets in networks like Downtown, the Financial District, and Chinatown may yield substantial resilience gains for certain modes.

\section{Discussion}

These findings suggest the influence of planning eras, design paradigms, transportation technologies, topography, and economics on US street network density, resilience, and connectedness. Overall, every large US metropolis is characterized by its preponderance of 3-way intersections. Sprawling suburban neighborhoods rank low on density and connectedness. The orthogonal grids we see in the downtowns of Portland and San Francisco have high density (i.e., intersection and street densities), connectedness (i.e., average number of streets per node), and order (based on circuity and statistical dispersion of node types), but low resilience in the presence of one-way streets, measured by MBC- and ANC-increases when switching from one-way to bidirectional streets.

A critical takeaway is that scale matters. The median average circuity is lower across the neighborhoods data set than across the municipal set, which in turn is lower than across the urbanized areas set. Conversely, the median average number of streets per node is higher across the neighborhoods data set than across the municipal set, which in turn is higher than across the urbanized areas set. The median intersection density per km\textsuperscript{2} is about 83\% higher in the neighborhoods data set than in the municipal or urbanized areas sets. These findings make sense: the Zillow neighborhood boundaries focus on large, core cities with older and denser street networks. The municipal boundaries only include incorporated cities and towns---discarding small census-designated places and unincorporated communities. The urbanized area boundaries include far-flung sprawling suburbs.

The characteristics of city street networks fundamentally depend on what \emph{city} means: municipal boundaries, urbanized areas, or certain central neighborhoods? The first is a legal/political definition, but captures the scope of city planning authority and decision-making for top-down interventions into human circulation. The second captures a wider self-organized human system and its emergent built form, but tends to aggregate multiple heterogeneous forms together into a single unit of analysis. The third captures the nature of the local built environment and lived experience, but at the expense of a broader view of the urban system and metropolitan-scale trip-taking. In short, multiple scales in concert provide planners and scholars a clearer view of the urban form and the topological and metric complexity of the street network than any single scale can.

This analysis finds a strong linear relationship, invariant across scales, between total street length and the number of nodes in a network. This differs from previous findings in the literature that relied on smaller sample sizes and examined European instead of US cities. We also find that most networks typically follow right-skewed distributions (particularly the exponentiated Weibull distribution) of street segment lengths. As discussed, this finding seems to make sense theoretically and is supported by these large-sample data at multiple scales, but obvious exceptions exist in those networks that exhibit substantial uniformity. At the neighborhood scale, examples include downtowns with consistent orthogonal grids, such as that of Portland, Oregon. At the municipal scale, examples include towns in the Great Plains that have orthogonal grids with consistent block sizes, platted at one time, and never subjected to expansion or sprawl.

These findings reveal urban form legacies of the practice and history of US planning. The spatial signatures of the Homestead Act, successive land use regulations, urban design paradigms, and planning instruments remain clearly etched in these cities' urban forms and street networks today. Accordingly when comparing median municipal street networks in each state, Nebraska has the lowest circuity, the highest average number of streets per node, the second shortest average street segment length, and the second highest intersection density. These findings illustrate how street networks across the Great Plains developed all at once, but grew very little afterwards---unlike, for instance, cities in California that were settled in the same era but later subjected to sprawl.

Future research could incorporate temporal analyses that go beyond the present study's cross-sectional data. This empirical analysis emphasized network structure, but further linking structural complexity to the temporal complexity of city dynamics and processes lies ahead as critical work. As OSMnx can automatically calculate several dozen street network measures, future work can use dimensionality reduction to identify significant baskets of indicators and cluster places into morphological types. These variables can also be used as advanced urban form measures in hedonic regressions and accessibility studies. Finally, future research can further explore urban spatial geometries such as block shapes and configurations, the statistical distributions of various indicators, and the comparative character of worldwide cities: this analysis of US urbanism and its specific empirical findings do not necessarily apply universally to cities elsewhere in the world.

\section{Conclusion}

This paper had three primary purposes. First, it presented empirical urban morphological findings from metric and topological analyses of the street networks of every US city/town, urbanized area, and Zillow neighborhood---particularly focusing on density, connectedness, and resilience. Second, its methods demonstrate the use of OSMnx as a new street network research toolkit, suggesting to urban planners and scholars new methods for acquiring and analyzing data consistently and at scale. Third, it built on past findings about the distribution of street segment lengths and the relationship between the total street length and the number of nodes in a network. This study has made all of these network datasets---for 497 urbanized areas, 19,655 cities and towns, and 6,857 neighborhoods---along with all of their attribute data and morphological measures available in an online public repository for other researchers to study and repurpose.

\section*{Acknowledgments}

The author wishes to thank Paul Waddell, Robert Cervero, David O'Sullivan, Elizabeth Macdonald, and Luis Bettencourt for their helpful comments and suggestions.

\IfFileExists{\jobname.ent}{\theendnotes}{}

\setlength{\bibsep}{0.00cm plus 0.05cm}
\bibliographystyle{apalike}
\bibliography{references}

\begin{thebibliography}{}

\bibitem[Agryzkov et~al., 2012]{agryzkov_algorithm_2012}
Agryzkov, T., Oliver, J.~L., Tortosa, L., and Vicent, J.~F. (2012).
\newblock An algorithm for ranking the nodes of an urban network based on the
  concept of {PageRank} vector.
\newblock {\em Applied Mathematics and Computation}, 219(4):2186--2193.

\bibitem[Albert and Barabási, 2002]{albert_statistical_2002}
Albert, R. and Barabási, A.-L. (2002).
\newblock Statistical mechanics of complex networks.
\newblock {\em Reviews of Modern Physics}, 74(1):47.

\bibitem[Albrecht and Abramovitz, 2014]{albrecht_indicator_2014}
Albrecht, J. and Abramovitz, M. (2014).
\newblock Indicator {Analysis} for {Unpacking} {Poverty} in {New} {York}
  {City}.
\newblock Technical report, CUNY, New York, NY.

\bibitem[Barthelemy, 2004]{barthelemy_betweenness_2004}
Barthelemy, M. (2004).
\newblock Betweenness centrality in large complex networks.
\newblock {\em The European Physical Journal B: Condensed Matter and Complex
  Systems}, 38(2):163--168.

\bibitem[Barthelemy, 2011]{barthelemy_spatial_2011}
Barthelemy, M. (2011).
\newblock Spatial networks.
\newblock {\em Physics Reports}, 499(1-3):1--101.

\bibitem[Barthelemy and Flammini, 2008]{barthelemy_modeling_2008}
Barthelemy, M. and Flammini, A. (2008).
\newblock Modeling {Urban} {Street} {Patterns}.
\newblock {\em Physical Review Letters}, 100(13).

\bibitem[Batty, 2005]{batty_network_2005}
Batty, M. (2005).
\newblock Network geography: {Relations}, interactions, scaling and spatial
  processes in {GIS}.
\newblock In Unwin, D.~J. and Fisher, P., editors, {\em Re-{Presenting} {GIS}},
  pages 149--170. John Wiley \& Sons, Chichester, England.

\bibitem[Beineke et~al., 2002]{beineke_average_2002}
Beineke, L.~W., Oellermann, O.~R., and Pippert, R.~E. (2002).
\newblock The {Average} {Connectivity} of a {Graph}.
\newblock {\em Discrete Mathematics}, 252(1):31--45.

\bibitem[Besbris et~al., 2015]{besbris_effect_2015}
Besbris, M., Faber, J.~W., Rich, P., and Sharkey, P. (2015).
\newblock Effect of neighborhood stigma on economic transactions.
\newblock {\em Proceedings of the National Academy of Sciences},
  112(16):4994--4998.

\bibitem[Boeing, 2017]{boeing_osmnx:_2017}
Boeing, G. (2017).
\newblock {OSMnx}: {New} {Methods} for {Acquiring}, {Constructing},
  {Analyzing}, and {Visualizing} {Complex} {Street} {Networks}.
\newblock {\em Computers, Environment and Urban Systems}, 65(126-139).

\bibitem[Boeing, 2018a]{boeing_morphology_2018}
Boeing, G. (2018a).
\newblock The {Morphology} and {Circuity} of {Walkable} and {Drivable} {Street}
  {Networks}.
\newblock In D'Acci, L., editor, {\em Mathematics of {Urban} {Morphology}
  (forthcoming)}. Birkhäuser, Cham, Switzerland.

\bibitem[Boeing, 2018b]{boeing_planarity_2018}
Boeing, G. (2018b).
\newblock Planarity and {Street} {Network} {Representation} in {Urban} {Form}
  {Analysis}.
\newblock arXiv:1806.01805:1--13.

\bibitem[Brandes and Erlebach, 2005]{brandes_network_2005}
Brandes, U. and Erlebach, T., editors (2005).
\newblock {\em Network analysis: methodological foundations}.
\newblock Number 3418 in Lecture {Notes} in {Computer} {Science}. Springer,
  Berlin, Germany.

\bibitem[Brin and Page, 1998]{brin_anatomy_1998}
Brin, S. and Page, L. (1998).
\newblock The anatomy of a large-scale hypertextual web search engine.
\newblock {\em Computer Networks and ISDN Systems: Proceedings of the Seventh
  International World Wide Web Conference}, 30(1-7):107--117.

\bibitem[Buhl et~al., 2006]{buhl_topological_2006}
Buhl, J., Gautrais, J., Reeves, N., Solé, R.~V., Valverde, S., Kuntz, P., and
  Theraulaz, G. (2006).
\newblock Topological patterns in street networks of self-organized urban
  settlements.
\newblock {\em The European Physical Journal B: Condensed Matter and Complex
  Systems}, 49(4):513--522.

\bibitem[Cardillo et~al., 2006]{cardillo_structural_2006}
Cardillo, A., Scellato, S., Latora, V., and Porta, S. (2006).
\newblock Structural properties of planar graphs of urban street patterns.
\newblock {\em Physical Review E}, 73(6).

\bibitem[Chin and Wen, 2015]{chin_geographically_2015}
Chin, W.-C.-B. and Wen, T.-H. (2015).
\newblock Geographically {Modified} {PageRank} {Algorithms}: {Identifying} the
  {Spatial} {Concentration} of {Human} {Movement} in a {Geospatial} {Network}.
\newblock {\em PLoS ONE}, 10(10):e0139509.

\bibitem[Corcoran et~al., 2013]{corcoran_analysing_2013}
Corcoran, P., Mooney, P., and Bertolotto, M. (2013).
\newblock Analysing the growth of {OpenStreetMap} networks.
\newblock {\em Spatial Statistics}, 3:21--32.

\bibitem[Costa et~al., 2007]{costa_characterization_2007}
Costa, L. d.~F., Rodrigues, F.~A., Travieso, G., and Villas~Boas, P.~R. (2007).
\newblock Characterization of complex networks: {A} survey of measurements.
\newblock {\em Advances in Physics}, 56(1):167--242.

\bibitem[Cranmer et~al., 2017]{cranmer_navigating_2017}
Cranmer, S.~J., Leifeld, P., McClurg, S.~D., and Rolfe, M. (2017).
\newblock Navigating the range of statistical tools for inferential network
  analysis.
\newblock {\em American Journal of Political Science}, 61(1):237--251.

\bibitem[Crucitti et~al., 2006a]{crucitti_centrality_2006-1}
Crucitti, P., Latora, V., and Porta, S. (2006a).
\newblock Centrality in networks of urban streets.
\newblock {\em Chaos: An Interdisciplinary Journal of Nonlinear Science},
  16(1):015113.

\bibitem[Crucitti et~al., 2006b]{crucitti_centrality_2006}
Crucitti, P., Latora, V., and Porta, S. (2006b).
\newblock Centrality measures in spatial networks of urban streets.
\newblock {\em Physical Review E}, 73(3):036125.

\bibitem[Dankelmann and Oellermann, 2003]{dankelmann_bounds_2003}
Dankelmann, P. and Oellermann, O.~R. (2003).
\newblock Bounds on the average connectivity of a graph.
\newblock {\em Discrete Applied Mathematics}, 129(2):305--318.

\bibitem[Dorogovtsev and Mendes, 2002]{dorogovtsev_evolution_2002}
Dorogovtsev, S. and Mendes, J. (2002).
\newblock Evolution of networks.
\newblock {\em Advances in Physics}, 51(4):1079--1187.

\bibitem[Ewing and Cervero, 2010]{ewing_travel_2010}
Ewing, R. and Cervero, R. (2010).
\newblock Travel and the {Built} {Environment}: {A} {Meta}-{Analysis}.
\newblock {\em Journal of the American Planning Association}, 76(3):265--294.

\bibitem[Frizzelle et~al., 2009]{frizzelle_importance_2009}
Frizzelle, B., Evenson, K., Rodriguez, D., and Laraia, B. (2009).
\newblock The importance of accurate road data for spatial applications in
  public health: customizing a road network.
\newblock {\em International Journal of Health Geographics}, 8(24).

\bibitem[Gleich, 2015]{gleich_pagerank_2015}
Gleich, D.~F. (2015).
\newblock {PageRank} {Beyond} the {Web}.
\newblock {\em SIAM Review}, 57(3):321--363.

\bibitem[Gudmundsson and Mohajeri, 2013]{gudmundsson_entropy_2013}
Gudmundsson, A. and Mohajeri, N. (2013).
\newblock Entropy and order in urban street networks.
\newblock {\em Scientific Reports}, 3.

\bibitem[Haklay, 2010]{haklay_how_2010}
Haklay, M. (2010).
\newblock How {Good} is {Volunteered} {Geographical} {Information}? {A}
  {Comparative} {Study} of {OpenStreetMap} and {Ordnance} {Survey} {Datasets}.
\newblock {\em Environment and Planning B: Planning and Design},
  37(4):682--703.

\bibitem[Huang et~al., 2016]{huang_trajgraph:_2016}
Huang, X., Zhao, Y., Ma, C., Yang, J., Ye, X., and Zhang, C. (2016).
\newblock {TrajGraph}: {A} {Graph}-{Based} {Visual} {Analytics} {Approach} to
  {Studying} {Urban} {Network} {Centralities} {Using} {Taxi} {Trajectory}
  {Data}.
\newblock {\em IEEE Transactions on Visualization and Computer Graphics},
  22(1):160--169.

\bibitem[Jiang, 2009]{jiang_ranking_2009}
Jiang, B. (2009).
\newblock Ranking spaces for predicting human movement in an urban environment.
\newblock {\em International Journal of Geographical Information Science},
  23(7):823--837.

\bibitem[Jiang and Claramunt, 2002]{jiang_integration_2002}
Jiang, B. and Claramunt, C. (2002).
\newblock Integration of space syntax into {GIS}: new perspectives for urban
  morphology.
\newblock {\em Transactions in GIS}, 6(3):295--309.

\bibitem[Jiang and Claramunt, 2004]{jiang_topological_2004}
Jiang, B. and Claramunt, C. (2004).
\newblock Topological {Analysis} of {Urban} {Street} {Networks}.
\newblock {\em Environment and Planning B: Planning and Design},
  31(1):151--162.

\bibitem[Jokar~Arsanjani et~al., 2015]{jokar_arsanjani_openstreetmap_2015}
Jokar~Arsanjani, J., Zipf, A., Mooney, P., and Helbich, M., editors (2015).
\newblock {\em {OpenStreetMap} in {GIScience}}.
\newblock Lecture {Notes} in {Geoinformation} and {Cartography}. Springer
  International, Cham, Switzerland.

\bibitem[Lee, 1979]{lee_kansas_1979}
Lee, L.~B. (1979).
\newblock {\em Kansas and the {Homestead} act, 1862-1905}.
\newblock The management of public lands in the {United} {States}. Ayer Company
  Publishers, North Stratford, NH.

\bibitem[Maron, 2015]{maron_how_2015}
Maron, M. (2015).
\newblock How complete is {OpenStreetMap}?
\newblock {\em Mapbox}.

\bibitem[Masucci et~al., 2009]{masucci_random_2009}
Masucci, A.~P., Smith, D., Crooks, A., and Batty, M. (2009).
\newblock Random planar graphs and the {London} street network.
\newblock {\em The European Physical Journal B: Condensed Matter and Complex
  Systems}, 71(2):259--271.

\bibitem[Newman, 2003]{newman_structure_2003}
Newman, M. E.~J. (2003).
\newblock The {Structure} and {Function} of {Complex} {Networks}.
\newblock {\em SIAM Review}, 45(2):167--256.

\bibitem[Newman, 2010]{newman_networks:_2010}
Newman, M. E.~J. (2010).
\newblock {\em Networks: {An} {Introduction}}.
\newblock Oxford University Press, Oxford, England.

\bibitem[Opsahl and Panzarasa, 2009]{opsahl_clustering_2009}
Opsahl, T. and Panzarasa, P. (2009).
\newblock Clustering in weighted networks.
\newblock {\em Social Networks}, 31(2):155--163.

\bibitem[O'Sullivan, 2014]{fischer_spatial_2014}
O'Sullivan, D. (2014).
\newblock Spatial {Network} {Analysis}.
\newblock In Fischer, M.~M. and Nijkamp, P., editors, {\em Handbook of
  {Regional} {Science}}, pages 1253--1273. Springer-Verlag, Berlin, Germany.

\bibitem[Over et~al., 2010]{over_generating_2010}
Over, M., Schilling, A., Neubauer, S., and Zipf, A. (2010).
\newblock Generating web-based 3d {City} {Models} from {OpenStreetMap}: {The}
  current situation in {Germany}.
\newblock {\em Computers, Environment and Urban Systems}, 34(6):496--507.

\bibitem[Porta et~al., 2006]{porta_network_2006}
Porta, S., Crucitti, P., and Latora, V. (2006).
\newblock The network analysis of urban streets: {A} dual approach.
\newblock {\em Physica A: Statistical Mechanics and its Applications},
  369(2):853--866.

\bibitem[Porterfield, 2005]{porterfield_homestead_2005}
Porterfield, J. (2005).
\newblock {\em The {Homestead} {Act} of 1862: {A} {Primary} {Source} {History}
  of the {Settlement} of the {American} {Heartland} in the {Late} 19th
  {Century}}.
\newblock Rosen Publishing Group, New York, NY, 1st edition.

\bibitem[Ratti, 2004]{ratti_space_2004}
Ratti, C. (2004).
\newblock Space syntax: some inconsistencies.
\newblock {\em Environment and Planning B: Planning and Design},
  31(4):487--499.

\bibitem[Schernthanner et~al., 2016]{schernthanner_spatial_2016}
Schernthanner, H., Asche, H., Gonschorek, J., and Scheele, L. (2016).
\newblock Spatial {Modeling} and {Geovisualization} of {Rental} {Prices} for
  {Real} {Estate} {Portals}.
\newblock In {\em {ICCSA} 2016}, pages 120--133. Springer, Cham, Switzerland.

\bibitem[Sherraden, 2005]{sherraden_inclusion_2005}
Sherraden, M. (2005).
\newblock {\em Inclusion in the {American} {Dream}: {Assets}, {Poverty}, and
  {Public} {Policy}}.
\newblock Oxford University Press, Oxford, England.

\bibitem[Southworth and Ben-Joseph, 1997]{southworth_streets_1997}
Southworth, M. and Ben-Joseph, E. (1997).
\newblock {\em Streets and the {Shaping} of {Towns} and {Cities}}.
\newblock McGraw-Hill, New York, NY.

\bibitem[Strano et~al., 2013]{strano_urban_2013}
Strano, E., Viana, M., da~Fontoura~Costa, L., Cardillo, A., Porta, S., and
  Latora, V. (2013).
\newblock Urban {Street} {Networks}, a {Comparative} {Analysis} of {Ten}
  {European} {Cities}.
\newblock {\em Environment and Planning B: Planning and Design},
  40(6):1071--1086.

\bibitem[Talen, 2003]{talen_measuring_2003}
Talen, E. (2003).
\newblock Measuring {Urbanism}: {Issues} in {Smart} {Growth} {Research}.
\newblock {\em Journal of Urban Design}, 8(3):195--215.

\bibitem[Trudeau, 1994]{trudeau_introduction_1994}
Trudeau, R.~J. (1994).
\newblock {\em Introduction to {Graph} {Theory}}.
\newblock Dover Publications, New York, NY, 2nd edition.

\bibitem[{U.S. Census Bureau}, 2010]{u.s._census_bureau_2010_2010}
{U.S. Census Bureau} (2010).
\newblock 2010 {Census} {Urban} and {Rural} {Classification} and {Urban} {Area}
  {Criteria}.

\bibitem[Viana et~al., 2013]{viana_simplicity_2013}
Viana, M.~P., Strano, E., Bordin, P., and Barthelemy, M. (2013).
\newblock The simplicity of planar networks.
\newblock {\em Scientific Reports}, 3(3495):1--6.

\bibitem[Willis, 2008]{willis_openstreetmap_2008}
Willis, N. (2008).
\newblock {OpenStreetMap} project completes import of {United} {States} {TIGER}
  data.
\newblock {\em Linux.com}.

\bibitem[Wu et~al., 2005]{wu_improving_2005}
Wu, J., Funk, T.~H., Lurmann, F.~W., and Winer, A.~M. (2005).
\newblock Improving spatial accuracy of roadway networks and geocoded
  addresses.
\newblock {\em Transactions in GIS}, 9(4):585--601.

\bibitem[Zhong et~al., 2017]{zhong_revealing_2017}
Zhong, C., Schläpfer, M., Arisona, S.~M., Batty, M., Ratti, C., and Schmitt,
  G. (2017).
\newblock Revealing centrality in the spatial structure of cities from human
  activity patterns.
\newblock {\em Urban Studies}, 54(2):437--455.

\bibitem[Zielstra and Hochmair, 2011]{zielstra_comparative_2011}
Zielstra, D. and Hochmair, H. (2011).
\newblock Comparative {Study} of {Pedestrian} {Accessibility} to {Transit}
  {Stations} {Using} {Free} and {Proprietary} {Network} {Data}.
\newblock {\em Transportation Research Record: Journal of the Transportation
  Research Board}, 2217:145--152.

\end{thebibliography}

\end{multicols}
\end{document}